# Physics Needs Philosophy. Philosophy Needs Physics

Carlo Rovelli[1]

**Abstract** Contrary to claims about the irrelevance of philosophy for science, I argue that philosophy has had, and still has, far more influence on physics than is commonly assumed. I maintain that the current anti-philosophical ideology has had damaging effects on the fertility of science. I also suggest that recent important empirical results, such as the detection of the Higgs particle and gravitational waves, and the failure to detect supersymmetry where many expected to find it, question the validity of certain philosophical assumptions common among theoretical physicists, inviting us to engage in a clearer philosophical reflection on scientific method.

## 1.

*Against Philosophy* is the title of a chapter of a book by one of the great physicists of the last generation: Steven Weinberg, Nobel Prize winner and one of the architects of the Standard Model of elementary particle physics [1]. Weinberg argues eloquently that philosophy is more damaging than helpful for physics—although it might provide some good ideas at times, it is often a straightjacket that physicists have to free themselves from. More radically, Stephen Hawking famously wrote that "philosophy is dead" because the big questions that used to be discussed by philosophers are now in the hands of physicists [2]. Similar views are widespread among scientists, and scientists do not keep them to themselves. Neil de Grasse Tyson, a well known figure in the popularisation of science in America, publicly stated in the same vein: "…we

✉ Carlo Rovelli
rovelli@cpt.univ-mrs.fr

[1] CPT, Aix-Marseille Université, Université de Toulon, CNRS, 13288 Marseille, France

learn about the expanding universe, … we learn about quantum physics, each of which falls so far out of what you can deduce from your armchair that the whole community of philosophers … was rendered essentially obsolete. [3]"

I disagree with these opinions. I present here some counter-arguments, arguing that philosophy has always played an essential role in the development of science, physics in particular, and is likely to continue to do so.

This is a long-standing debate. In this regard, a delightful chapter was played out in Athens during its classical period. At the time, the golden youth of the city was educated in famous schools. Among these, two stood out: the school of Isocrates, and the *Academy,* founded by a certain Plato. The rivalry between the two was heated, like the rivalry between Oxford and Cambridge, but was not just about quality: the very approach to education was different in the two schools. Isocrates offered a high-level *practical* education, teaching the youth of Athens the kind of skills and knowledge required to become politicians, lawyers, judges, architects, and so on. The Academy on the other hand focused rather on discussing *general questions* about foundations: What is justice? What would be the best laws? What is beauty? What is matter made of? And Plato had invented a good name for this way of posing problems: "philosophy".

The disagreement between the two schools was deep-rooted. Isocrates' criticisms of Plato's approach to education and knowledge were direct:

> Those who do philosophy, who determine the proofs and the arguments … and are accustomed to enquiring, but take part in none of their practical functions, … even if they happen to be capable of handling something, they automatically do it worse, whereas those who have no knowledge of the arguments [of philosophy], if they are trained [in concrete sciences] and have correct opinions, are altogether superior for all practical purposes. Hence for sciences, philosophy is entirely useless [4].

This is remarkably like the claim by those contemporary scientists who argue that philosophy has no role to play in science.

As it happened, a brilliant young student in Plato's school wrote a short work in response to Isocrates' criticisms. This was the *Protrepticus*, a text that became famous in antiquity. It has survived only in part, and we know it only through reconstruction from extensive quotations by later writers. A group of classical scholars led by Doug Hutchinson and Monte Ransome Jonson have recently completed a novel reconstruction which can be accessed online [5]. The Protrepticus was probably written in the form of a dialog between characters defending opposing positions—dialog was of course the style preferred by Plato. What remains of the text is sufficient to understand the main arguments this young student put forward in his reply to Isocrates, in defence of philosophy.

The bright young fellow who authored the pamphlet later left Athens, but eventually returned to open his own school, and had quite a career. His name was Aristotle.

Two millennia of development of the sciences and philosophy have vindicated and, if anything, strengthened Aristotle's defence of philosophy against Isocrates' accusations of futility. His arguments are still remarkably relevant, and I take inspiration from them to respond here to the *current* claims that philosophy is useless to physics.

**2.**

The first of Aristotle's arguments is the fact that

> general theory supports and happens to be useful for the development of practice.

Let me separate two aspects of the issue: first, the relevance of philosophy to science in the past, and second, the question of whether philosophy has become irrelevant for science today. Let us start with the first.

Today, after a couple of millennia, during which both philosophy and science have developed considerably, historical evidence regarding the influence of philosophy on science has become overwhelming.

Here are some instances of this influence, drawn from astronomy and physics. I shall only mention a few examples. Ancient astronomy—that is, everything we know about the Earth being round, its size, the size of the moon and the sun, the distances to the moon and the sun, the motion of the planets in the sky, and the basis from which modern astronomy and modern physics have emerged—, is a direct descendent of philosophy. The key questions that motivated these developments were posed in the *Academy* and the *Lyceum,* and motivated by highly theoretical, rather than practical concerns. Although centuries later Galileo and Newton took great steps beyond previous physics and astronomy, and in particular, the Aristotelian world view, they relied heavily on what had come before [6]. Galileo and Newton extended previous knowledge, reinterpreting, reframing, and building upon it. Galileo's understanding in particular would have been inconceivable without Aristotelian physics.

More importantly, Galileo's work would have been inconceivable without the ideology he derived from Plato, namely, to search for the ideal mathematical order underlying appearances. Galileo was guided by an almost fanatic Platonism. In his work, Newton was explicit about his debt to ancient philosophy, Democritus in particular, for ideas that arose originally from philosophical motivations, such as the notions of empty space, atomism, and natural rectilinear motion. Furthermore, his crucial discussion about the nature of space and time built upon his discussions with (and against) Descartes.

But the direct influence of philosophy on physics is certainly not limited to the birth of modern physics. It can be recognised in every major step. Take the twentieth century. Both major advances made by twentieth century physics were strongly influenced by philosophy. They would have been inconceivable without the philosophy of the time. Quantum mechanics springs from an intuition due to Heisenberg, grounded in the strongly positivist philosophical atmosphere in which he found himself: one gets knowledge by restricting oneself to what is observable. The abstract of Heisenberg's 1925 milestone paper on quantum theory is explicit about this:

> The aim of this work is to set the basis for a theory of quantum mechanics based exclusively on relations between quantities that are in principle observable [7].

The same distinctly philosophical attitude nourished Einstein's discovery of special relativity: by restricting to what is observable, we recognise that the notion of simultaneity is misleading. Einstein explicitly recognised his debt to the philosophical writings of Mach and Poincaré. Without these inputs, his special relativity would have been inconceivable. Although not the same, the philosophical influences on Einstein's conception of general relativity were even stronger. Once again, he was explicit in recognising his debt to philosophy, this time to the critical thinking of Leibniz, Berkeley, and Mach. Einstein's relation with philosophy was indeed complex: he claimed for instance that Schopenhauer had had a pervasive influence on him. This is less evident to detect in his physics, but Schopenhauer's ideas on time and representation are perhaps not so hard to recognise in Einstein's ideas leading to general relativity; this influence has also been studied [8]. Can it really be a coincidence that, in his younger days, the greatest physicist of the twentieth century should have had such a clear focus on philosophy [9], reading Kant's three *Critics* when he was 15 years old?

Why this influence? Because philosophy can provide methods for producing new ideas, novel perspectives, and critical thinking. Philosophers have tools and skills that physics needs, but do not belong to the physicists training: conceptual analysis, attention to ambiguity, accuracy of expression, the ability to detect gaps in standard arguments, to devise radically new perspectives, to spot conceptual weak points, and to seek out alternative conceptual explanations. Nobody puts this better than Einstein himself:

> A knowledge of the historic and philosophical background gives that kind of independence from prejudices of his generation from which most scientists are suffering. This independence created by philosophical insight is—in my opinion—the mark of distinction between a mere artisan or specialist and a real seeker after truth [10].

Put more strongly, it is sometimes said that: "Scientists do not do anything unless they first get permission from philosophy."

Thus, if we read what the greatest scientists had to say about the usefulness of philosophy, physicists like Heisenberg, Schrödinger, Bohr, and Einstein, we find that they expressed totally opposite opinions to those of Hawking and Weinberg.

### 3.

Hers is a second argument due to Aristotle:

> Those who deny the utility of philosophy, are doing philosophy.

This point is less trivial than it sounds at first. Let's examine what Weinberg and Hawking write. Both have obtained important scientific results. Weinberg, for instance, found the correct symmetry group to describe the interactions between the elementary particles, while Hawking discovered that black holes are hot, and computed their temperature. In doing this kind of thing, they were doing science. In writing things like "philosophy is useless to physics", or "philosophy is dead", they were not doing physics. So what were they doing? They were reflecting on the best way to develop science.

The issue here is the *methodology* of science. A central concern in the philosophy of science is of course precisely to ask *how* science is done and how it *could* be done to be more effective. Good scientists *do* reflect on their own methodology, and it is quite appropriate that Weinberg and Hawking have done so too. But how?

They express a *certain* idea about the methodology of science. Is this the eternal truth about how science has always worked and should work? Is it the best understanding of science we have at present?

It is neither. In fact, it is not difficult to trace the origins of this idea. It arises from the background of logical positivism, corrected by Popper and Kuhn. The current dominant methodological ideology in theoretical physics derives from their notions of *falsifiability* and *scientific revolution*, which are popular among theoretical physicists; they are often mentioned and are commonly used to orient research and evaluate scientific work.

Therefore, in declaring the uselessness of philosophy, Weinberg, Hawking, and other "anti-philosophical" scientists are in fact paying homage to the *philosophers* of science they have read, or whose ideas they have absorbed from their environment. The imprint is unmistakable. When viewed as an ensemble of pseudo-statements, words that resemble statements but have no proper meaning, of the kind recurrent for instance in the way Neil de Grasse Tyson mocks philosophy, these criticisms are easily traced to the Vienna Circle's anti-metaphysical stance [11]. Behind these anathemas against "philosophy", one can almost hear the Vienna Circle's slogan of "no metaphysics!"

Thus, when Weinberg and Hawking state that philosophy is useless, they are actually stating their adhesion to a particular philosophy of science. In principle, there's nothing wrong with that; but the problem is that it is *not* a very good philosophy of science.

On the one hand Newton, Maxwell, Boltzmann, Darwin, Lavoisier, and so many other major scientists worked within a very different methodological perspective, and did pretty good science as well. On the other hand, philosophy of science has advanced since Carnap, Popper, or Kuhn, recognising that the way science effectively works is richer and more subtle than the way it was portrayed in the analysis of these thinkers. Weinberg and Hawking's error is to mistake a particular, historically circumscribed, limited understanding of science for something like the eternal logic of science itself. The weakness of their position is the lack of awareness of its frail historical contingency. They present science as a discipline with an obvious and uncontroversial methodology, as if this had been the same from Bacon to the detection of gravitational waves, or as if it was completely obvious *what* we should be doing and *how* we should be doing it when we do science.

The reality is quite different. Science has repeatedly redefined its own understanding of itself, along with its goals, its methods, and its tools. This flexibility has played a major role in its success. Let us consider a few examples from physics and astronomy. In light of Hipparchus and Ptolemy's extraordinarily successful predictive theories, the proper goal of astronomy was to find the right combination of circles to describe the *motion of the heavenly bodies around the Earth*. But contrary to expectations, it turned out that Earth was itself one of the heavenly bodies. And indeed after Copernicus, the proper goal appeared to be to find the right combination of *moving spheres* that would reproduce the motion of the planets around the Sun. But once again contrary to expectations, it turned out that abstract elliptical trajectories were better than spheres. Then after Newton, it seemed clear that the aim of physics was to find the *forces acting on bodies*. Contrary to this program, it turned out that the world could be better described by dynamical fields rather than bodies. After Faraday and Maxwell, it was clear that physics had to find laws of *motion in space, as time passes*. Contrary to assumptions, it turned out that space and time are themselves dynamical. After Einstein, it finally became clear that physics must only search for the *deterministic laws of Nature*. Contrary to expectations, it turned out that we can at best give probabilistic laws. And so on. Here are some sliding definitions for what scientists have thought science to be: deduction of general laws from observed phenomena, finding out the ultimate constituents of Nature, accounting for regularities in empirical observations, finding provisional conceptual schemes for making sense of the world. (The last one is the one I like.)

Science is not a project with a methodology written in stone, well circumscribed objectives, or a fixed conceptual structure. It is our ever evolving endeavour to better understand the world. In the course of its development, it has repeatedly violated its own rules and its own stated methodological assumptions.

A currently common description of what scientists do—the one we learn today at school—is collecting data (observations, experiments, measurements) and making sense of them in the form of theories. The relationship between data and theory is complex, and far from being uncontroversial, since it it not at all obvious how we get from data to theory, and it is not obvious how are theory-laden are the data themselves. But let's gloss over that. As time goes by, new data are acquired and theories evolve. In this picture scientists are depicted as rational beings who play this game using their intelligence, a specific language, and a well established cultural and conceptual structure.

The problem with this picture is that conceptual structures evolve as well. Science is not simply an increasing body of empirical information we have about the world and a sequence of changing theories. It is also the evolution of our own conceptual structure. It is the continuous search for the best conceptual structure for grasping the world, at a given level of knowledge. And the modification of the conceptual structure needs to be achieved from within our own thinking, rather as a sailor must rebuild his own boat while sailing, to use the beautiful simile of Otto Neurath so often quoted by Quine [12].

This intertwining of learning and conceptual change, this flexibility, and this evolution of methodology and objectives, have developed historically in a constant dialogue between practical science and philosophical reflection. This is a further reason why so much science has been deeply influenced by philosophical reflection. The views of scientists, whether they like it or not, are impregnated with philosophy.

And here we come back to Aristotle: "Philosophy provides guidance how research must be done."

Not because philosophy can offer a final word about the right methodology of science (contrary to the *philosophical* stance of Weinberg and Hawking). But because philosophers have conceptual tools for addressing the issues raised by this continuous conceptual shift. The scientists that deny the role of philosophy in the advancement of science are those who think they have already found the final methodology, that is, that they have already exhausted and answered all methodological questions. They are consequently less open to the conceptual flexibility needed to go ahead. They are the ones trapped in the ideology of their time.

**4.**

I think an even stronger case can be made. I believe that one reason for the relative sterility of theoretical physics over the last few decades may well be *precisely* that the wrong philosophy of science is held dear today by many physicists. Popper and Kuhn, so popular among theoretical physicists, have shed light on important aspects of the way good science works, but their picture of science is incomplete and I suspect that, taken prescriptively and uncritically, their insights have ended up misleading research. Let's see why.

Kuhn's emphasis on discontinuity and incommensurability has misled many theoretical and experimental physicists into disvaluing the formidable *cumulative* aspects of scientific knowledge. Popper's emphasis on falsifiability, originally a demarcation criterion, has been flatly misinterpreted as an evaluation criterion. The combination of the two has given rise to disastrous methodological confusion: the idea that past knowledge is irrelevant when searching for new theories, that all unproven ideas are equally interesting and all unmeasured effects are equally likely to occur, and that the work of a theoretician consists in pulling arbitrary possibilities out of the blue and developing them, since anything that has not yet been falsified might in fact be right.

This is the current "why not?" ideology: any new idea deserves to be studied, just because it has not yet been falsified; any idea is equally probable, because a step further ahead on the knowledge trail there may be a Kuhnian discontinuity that was not predictable on the basis of past knowledge; any experiment is equally interesting, provided it tests something as yet untested.

I think that this methodological philosophy has given rise to mountains of useless theoretical work in physics and many useless experimental investments.

Arbitrary jumps in the unbounded space of possibilities have *never* been an effective way to do science. The reason is twofold: first, there are too many possibilities, and the probability of stumbling on a good one by pure chance is negligible; but more importantly, nature always surprises us and we, the limited critters that we are, are far less creative and imaginative than we may think. When we consider ourselves to be "speculating widely", we are mostly playing out rearrangements of old tunes: true novelty that works is not something we can just find by guesswork.

The most radical conceptual shifts and the most unconventional ideas that have actually worked have in fact always been strictly motivated, almost forced, either by the overwhelming weight of new data, or by a well-informed analysis of the internal contradictions within existing, successful theories. Science works through continuity, not discontinuity.

Examples of the first case—novelty forced by data—are Kepler's ellipses, and quantum theory. Kepler did not just "come out with the idea" of ellipses: nature had to splash ellipses on his face before he could see them. He was using ellipses as an approximation for the deferent-epicycle motion of Mars and was astonished to find that the approximation worked better than his model [13]. Similarly, atomic physicists of the early twentieth century struggled long and hard against the idea of discontinuities in the basic laws, doing everything they could to avoid accepting the clear message from spectroscopy, that is, that there was actually discontinuity in the very heart of mechanics. In both instances, the important new idea was forced by data.

Examples of the second case—radical novelty from old theories—are the heliocentric system and general relativity. Neither Copernicus nor Einstein relied significantly on new data. But neither did their ideas come out of the blue. They both started from an insightful analysis of successful well-established theories: Ptolemaic astronomy, Newtonian gravity, and special relativity. The contradictions and unexplained coincidences they found in these would open the way to a new conceptualisation.

In any case, it is not merely fishing out unfalsified theories and testing them that brings results. Rather, it is a sophisticated use of induction, building upon a vast and ever growing *accumulation* of empirical and theoretical knowledge, that provides the hints we need to move ahead. It is by focusing on empirically successful insights that we move ahead. Einstein's "relativity" was not a "new idea": it was Einstein's realisation of the extensive validity of Galilean relativity. There was no discontinuity: in fact it was continuity at its best. It was Einstein's insightful "conservatism" in the face of those who were too ready to discard the relativity of velocity just because of Maxwell's equations.

I think this lesson is missed by much contemporary theoretical physics, where plenty of research directions are too quick to discard what we have already found out about Nature.

Ironically, indeed, the recent momentous steps taken by experimental physics are all rebuttals of today's freely speculative attitude toward theoretical physics. Three major empirical results have marked recent fundamental physics: gravitational waves, the Higgs, and the absence of supersymmetry at LHC. All three are confirmations of old physics and disconfirmations of widespread speculation. In all three cases, Nature is telling us: do not speculate so freely. So let's look more closely at these examples.

The detection of gravitational waves, rewarded by the last Nobel Prize in fundamental physics, has been a radical confirmation of century-old general relativity. But it is more than this. The recent nearly simultaneous detection of gravitational and electromagnetic signals from the merging of two neutron stars (the event called GW170817) has improved our knowledge of the ratio between the speeds of propagation of gravity and electromagnetism by something like 14 orders of magnitudes in a single stroke [14, 15]. One consequence of this momentous increase in our empirical knowledge has been to rule out a great many theories put forward as alternatives to general relativity, ideas that have been studied by a large community of theoreticians over the last few decades, confirming instead the century-old general relativity as the best theory of gravity available at present.

The well publicised detection of the Higgs particle at CERN has confirmed the Standard Model of particle physics (founded by Steven Weinberg, among others) as the best current theory for high energy physics, against scores of later alternatives that have long been receiving much attention.

CERN's emphasis on the discovery of the Higgs when the Large Hadron Collider became operational in Geneva has also served to hide the true surprise that emerged from this particular exploration of high energy physics: the *absence* of supersymmetric particles where a generation of theoretical physicists had been expecting to find them. Despite the rivers of ink and flights of fancy, the minimal supersymmetric standard model suddenly finds itself in deep trouble. So once again, Nature has seriously rebuffed the free speculations of a large community of theoretical physicists who ended up believing them.

I think that Nature's repeated snub of the current methodology in theoretical physics should encourage a certain humility, rather than arrogance, in our philosophical attitude.

I suspect that part of the problem is precisely that the dominant ideas of Popper and Kuhn have misled current theoretical investigations. Physicists have been too casual in dismissing the insights of successful established theories. Misled by Kuhn's insistence on incommensurability across scientific revolutions, they fail to build on what we already know, which is how science has always moved forward. A good example of this is the disregard for general relativity's background independence in many attempts to incorporate gravity into the rest of fundamental physics.

Furthermore, the emphasis on falsifiability has made physicists blind to a fundamental aspect of scientific knowledge: the fact that credibility *has degrees* and that reliability can be extremely high, even when it is not absolute certainty. This has a doubly negative effect: considering the insights of successful theories as irrelevant for progress in science (because "they could be falsified tomorrow"), and failing to see that a given investigation may have little plausibility even if it has not yet been falsified.

The scientific enterprise is founded on degrees of credibility, which are constantly updated on the basis of new data or new theoretical developments. Recent attention to Bayesian accounts of confirmation in science is common in the philosophy of science, but largely ignored in the theoretical physics community, with negative effects, in my opinion [16].[1]

What I intend here is not a criticism of Popper and Kuhn, whose writings are articulate and insightful. What I am pointing out is that a simple-minded version of their outlooks has been taken too casually by many physicists as the ultimate word on the methodology of science.

Far from being immune from philosophy, current physics is deeply affected by philosophy. But the lack of philosophical awareness needed to recognise this influence, and the refusal to listen to philosophers who try to make amends for it, is a source of weakness for physics.

## 5.

Here is one last argument from the Protrepticus:

> More in need of philosophy are the sciences "where perplexities are greater".

Today fundamental physics *is* in a phase of deep conceptual change, because of the success of general relativity and quantum mechanics and the open "crisis" (in the sense of Kuhn, I would rather say "opportunity") generated by the current lack of an accepted quantum theory of gravity. This is why some scientists, including myself, working as I do on quantum gravity, are more acutely aware of the importance of philosophy for physics.

---

[1] The worst episode of this misunderstanding is the confusion between the the (strong) common-sense notion of 'confirmation' and the (weak) Bayesian notion of 'confirmation' that has driven the controversy over Richard Dawid's work on non-empirical confirmation [16]. An attempt to study the actual source of (possibly unjustified) confidence in a theory has been re-trumpeted by scientists as a proof of validity.

Here is a list of topics currently discussed in theoretical physics: What is space? What is time? What is the "present"? Is the world deterministic? Do we need to take the observer into account to describe nature? Is physics better formulated in terms of a "reality" or in terms of "what we observe", or is there a third option? What is the quantum wave function? What exactly does "emergence" mean? Does a theory of the totality of the universe make sense? Does it make sense to think that physical laws themselves might evolve? It is clear to me that input from past and current philosophical thinking cannot be disregarded in addressing these topics.

In loop quantum gravity, my own technical area, Newtonian space and time are reinterpreted as a manifestation of something which is granular, probabilistic, and fluctuating in a quantum sense. Space, time, particles, and fields get fused into a single entity: a quantum field that does not live in space or time. The variables of this field acquire definiteness only in interactions between subsystems. The fundamental equations of the theory have no explicit space or time variables. Geometry appears only in approximations. Objects exist within approximations. Realism is tempered by a strong dose of relationalism. I think we physicists need to discuss with philosophers, because I think we need help in making sense of all this.

**6.**

To close, I would like to add a brief word on the opposite issue: the relevance of science for philosophy.

I do so only because some manifestations of anti-philosophical attitudes in scientific circles are only a reaction to anti-scientific attitudes in some areas of philosophy and other humanities.

In the post-Heideggerian atmosphere that dominates some philosophy departments on the 'continent', ignorance of science is something to exhibit with pride. Science is not "authentic" knowledge; it misses true knowledge. "The botanist's plants are not the flowers of the hedgerow; the 'source' which the geographer establishes for a river is not the 'springhead in the dale [17]'" implying in the context that the only thing that counts is the latter.

And here is an example from another sector of the today's intellectual world, sociology: "…there is no obligation upon anyone framing a view of the world to take account of what twentieth-century science has to say [18] ." A comment which is either trivial ("there is no obligation to be intelligent") or misleading, in the etymological sense of leading in the wrong direction.

It seems clear to me that, just as the best science listens keenly to philosophy, so the best philosophy will listen keenly to science. This has certainly been so in the past: from Aristotle and Plato, to Descartes and Hume, Kant and Hegel, Husserl and Lewis, the best philosophy has always been closely tuned into science. No great philosopher of the past would ever have thought for a moment of not taking seriously the knowledge of the world offered by the science of their times.

Science is an integral and essential part of our culture. It is far from being capable of answering all the questions we would like to ask, but it is nevertheless an extremely powerful tool, able to address innumerable problems, including those that concern ourselves and the Universe as a whole. Our general knowledge is the result of the contributions from vastly different domains, from science to philosophy, all the way to literature and the arts, and our capacity to integrate them. Those philosophers that discount science, and there are many of them, do a serious disservice to intelligence and civilisation, in my opinion. When they say that entire fields of knowledge are impermeable to science, and that they are the ones who know better, they remind me of two little old men on a park bench: "Aaaah, "says one, his voice shaking," all these scientists who claim they can study consciousness, or the beginning of the universe." "Ohh," says the other, "how absurd! Of course they can't understand these things. But we do!"